\input{epsf}
\documentstyle[11pt,newpasp]{article}

\def\etal{{\it et al. }}

\def\etal{{\it et al. }}
\def\gsim{ \lower .75ex \hbox{$\sim$} \llap{\raise .27ex \hbox{$>$}} }
\def\lsim{ \lower .75ex \hbox{$\sim$} \llap{\raise .27ex \hbox{$<$}} }
\def\pp{\noindent\parshape 2 0truecm 12.0truecm 1.5truecm 10truecm}

\begin{document}

\title{Dynamical effects on galaxies in clusters}

\author{Ben Moore, Vicent Quilis \& Richard Bower}

\affil{Department of Physics, Durham University, South
Road, Durham City, DH1 3LE, UK} 

\begin{abstract}
The observed morphological evolution since $z\sim 0.5$ within galaxy clusters
provides evidence for a hierarchical universe. This evolution is driven
by dynamical effects that operate within the cluster environment -- suppression
of star-formation by ram-pressure and viscous stripping of the intra-galactic
medium and tidal heating of disks by gravitational encounters.  
\end{abstract}


\section{Evidence for a hierarchical universe?}

A hierarchical universe with a low matter density is the currently favoured model
for structure formation. However, observational evidence for evolution is
scarce - structures appear old and similar over moderate look back times 
(Peebles 1999).
Galaxy clusters are the most massive virialised objects in the universe and
are the latest systems to form, therefore they may reveal evidence for 
recent evolution.
Indeed, perhaps the strongest indication for evolution can be found
by examining galaxy morphologies in rich clusters over the past 5 Gyrs
({\it i.e.} references within: Sandage \etal 1970
Butcher \& Oemler 1978, 
Dressler 1980, 
Dressler \& Gunn 1983, 
Couch \etal 1987,  
Binggeli \etal 1988,
Ferguson \& Sandage 1991, 
Binggeli \etal 1991, 
Thompson \& Gregory 1993,
Peterson \& Caldwell 1993,
Lavery \& Henry 1994,
Barger \etal 1996,
Dressler \etal 1997, 
Balogh \etal 1998, 
Couch \etal 1998, 
Poggianti \etal 1999,
Bower \etal 1999).
At $z\sim 0.5$ we see that clusters of galaxies contain predominantly
late type disk galaxies that have undergone a transformation into dwarf
ellipticals (dSph) by the present day.  There is also compelling evidence that
the fraction of S0 galaxies decreases over the same look back time (Dressler
\etal 1997), whilst at even earlier epochs, tentative evidence for the
hierarchical growth of the cluster elliptical galaxies may be found (van Dokkum
\etal 1999).

The data suggest that the cluster environment is causing a morphological
transformation between galaxy types. Other indicators of environmental
influences in clusters include the morphology-density relation (Dressler
1980), signatures of star-bursts followed by a rapid truncation of
star-formation in cluster spirals and k+a/a+k galaxies (Poggianti \etal 1999),
the absence of low surface brightness disks in clusters (Bothun \etal 1993)
and the presence of a large component of diffuse star-light (Bernstein \etal
1993, Freeman - this volume).  In this paper we will discuss the role
of environment and the proposed mechanisms that are responsible for driving
galactic evolution in clusters and creating the ``morphological and
spectroscopic Butcher-Oemler effects''.


\section{Mergers, Winds, Harassment \& Stripping}

Several mechanisms have been proposed that can induce a morphological
transformation between galaxy classes. Toomre \& Toomre's (1972) pioneering
n-body simulations of merging spirals gave rise to elliptical remnants.  This
process is most effective when the encounter velocity is comparable to the
galaxies internal velocity, therefore mergers should be rare within 
rich virialised systems, a fact pointed out by many authors
({\it e.g.} Mamon - this volume), however see van Dokkum (1999).
Numerical simulations of clusters that formed naturally in a hierarchical
universe show that mergers are indeed rare (Ghigna \etal 1998).  Galactic
winds from supernovae are frequently cited as important mechanisms for dwarf
galaxy evolution (Dekel \& Silk 1986) yet observational evidence for this
process is not compelling (Martin 1998) and it is unlikely that feedback can
reshape the stellar configuration.

We therefore focus on the two mechanisms that are most likely to operate
extensively in the cluster environment ({\it c.f.} Fujita 1999).  Impulsive
heating via rapid tidal encounters has been demonstrated to be an efficient
mechanism at heating disk galaxies in clusters -- termed ``galaxy harassment''
by Moore \etal (1996, 1998), tidal forces and internal 
cluster dynamics have been studied by many authors, including
Richstone \& Malmuth 1983, Icke 1985, Merritt 1985, Byrd \&
Valtonen 1990, Valluri \& Jog 1991, Valluri 1993, Gnedin (1999).  
Gunn \& Gott (1978)
suggested that the ram-pressure force of the hot intra-cluster medium could
remove the atomic hydrogen from disks infalling into clusters. Nulsen (1982)
proposed that the viscous and turbulent stripping of gas would be equally
as effective.


Morphological transformation by gravitational encounters is a process that
requires several strong or frequent tidal collisions and acts over a timescale
of several $\times 10^9$ years. In contrast, hydrodynamic effects operate on a
much shorter timescale, $\sim 10^7$ years.  Observational evidence for
gas-dynamical interactions in clusters is scarce due to the speed at which the
process occurs (Stevens \etal 1999), however the literature does contain some
examples ({\it i.e.} Warmels 1988, Cayatte \etal 1990, Gavazzi \etal 1995,
Kenny \& Koopman 1999, and several recent studies in this volume by Vollmer
\etal (1999), Ryder
\etal (1997) \& J. Solanes \etal).

The first 3-dimensional calculation of ram pressure stripping using realistic
disk galaxy models was carried out by Abadi \etal (1999). Earlier work had
studied the properties of spherical systems in two dimensions ({\it i.e.}
Balsara \etal 1994).  Abadi \etal found that the simple analytic argument
proposed by Gunn \& Gott (1972) balancing the ram pressure, $\rho_{_{ICM}} v^2$, with
the disk's gravitational restoring force works very well. Simulations at
$45^\circ$ and edge on to the motion through the intra-cluster medium suffered
slightly less stripping than face on.  However, even in the best case scenario
for stripping, an $L_*$ disk galaxy moving through the center of the Coma
cluster at 3000 km s$^{-1}$, a significant fraction of gas remained 
(HI extending to a radius $\sim 5$ kpc) that would continue to form stars.

\

\centerline{\epsfysize=5.0truein \epsfbox{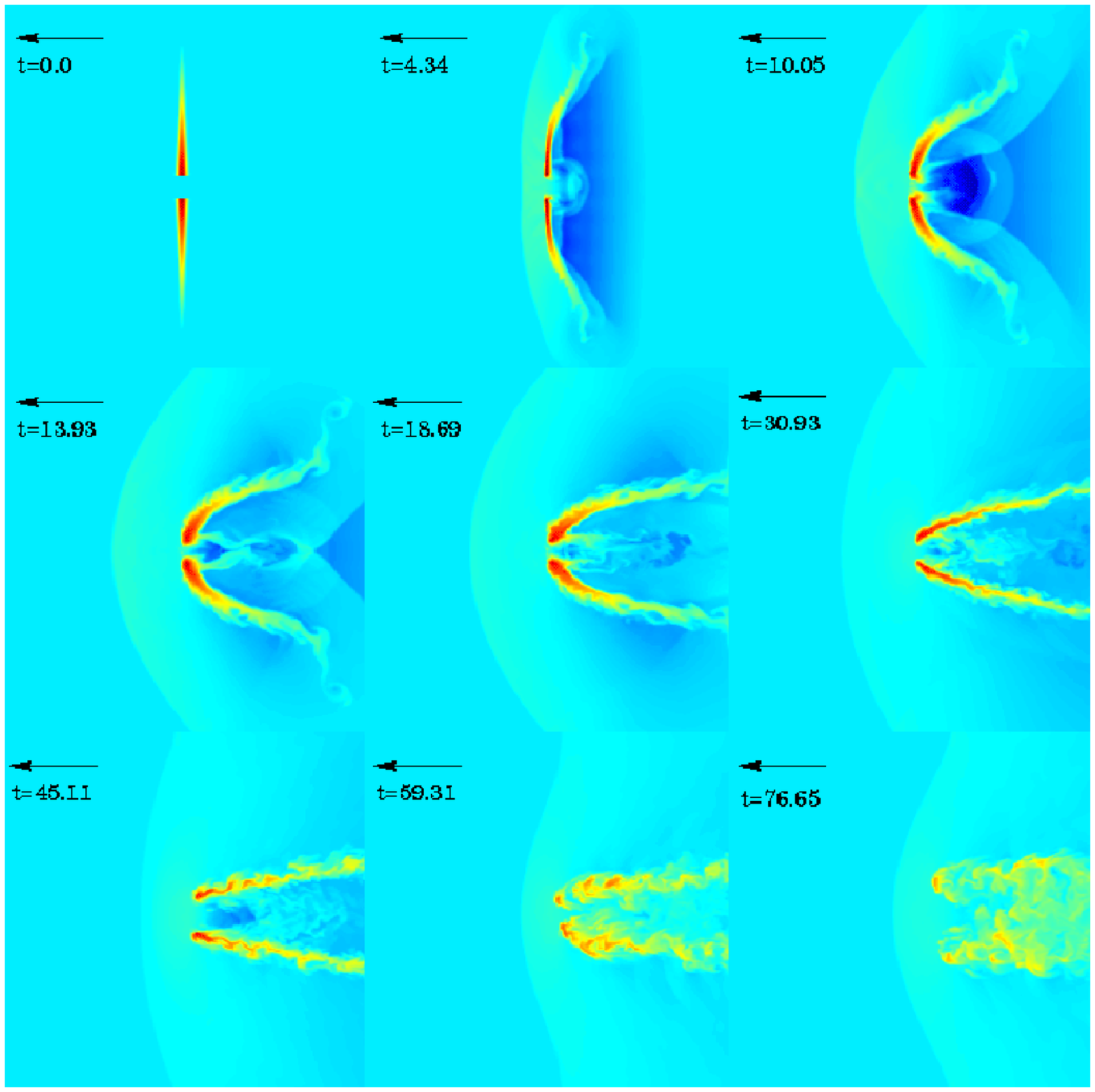}}

\

\noindent{\small {\bf Figure 1} \ \ The time evolution of the diffuse HI 
component within a spiral galaxy falling into the Coma cluster (Quilis \etal 1999).}

\

A more realistic treatment of the stripping process using a Eulerian finite
difference code revealed that the turbulent and viscous stripping processes
help to remove more gas than just the ram pressure alone (Quilis \etal 1999).
Quilis \etal also realised that disks are not smooth homogeneous mediums, but
have a complex structure containing many holes and regions that are completely
devoid of gas (Brinks \& Bajaja 1986).
Once these are included, the ICM streams 
through the holes, preventing
``backside infall'' and ablates away the gas from their edges,
rapidly removing the entire HI content of disks.  Thus, these authors find
that the stripping process alone, is in general 100\% efficient at removing
the HI component within a timescale $\lsim 10^7$ years (see Figure 1).

Without fresh material to replenish GMC's, star formation will be rapidly
truncated (Elmgreen \& Efremov 1997). 
Continued heating by tidal encounters with massive galaxies
increases the disk scale height by factors of 2--3. Spiral features will be
suppressed and these galaxies will resemble the S0 galaxies observed in nearby
clusters. The efficiency of the stripping process increases rapidly towards
the cluster center where $\rho_{_{ICM}}$ and $v$ are maximum, thus we have a natural way
of creating the morphology-density relation. Whether or not the transformation
of Sa/Sb galaxies to S0's can reproduce the same scaling as observed
remains to be tested.

Galaxy harassment will transform dI/Sc/Sd disks into dwarf elliptical (dSph)
galaxies over a timescale of several billion years - {\it i.e.} several
cluster orbital periods. Simulations of this process appear to produce remnant
galaxies that closely match the observed cluster dwarf's (Moore \etal 1996,
1998), although the sample of kinematical data is small.  A fundamental
prediction of the formation of dE (dSph) galaxies by tidal heating is that the
remnant galaxies are embedded in diffuse streams of stars, tidally removed
from the progenitor disks (see Figure 2). The surface brightness profiles of these galaxies
are expected to be well fitted by an exponential profile over their central
regions but a significant excess of stars will be found at 4--5 scale lengths
(Moore \etal 1999).  The surface brightness of this excess will occur fainter than
$27-28M_{_B}$ per arcsec$^{-2}$, an observable effect with 4m or 8m class
telescopes. The presence of freely orbiting planetary nebulae
in clusters (Freeman, this meeting) 
provides strong support for the efficiency of this process.

Not surprisingly, low surface brightness disks are not found within 
clusters (Bothun \etal 1993). 
The shear extent of their stellar distributions and low central
potentials make them unstable to tidal forces. Once these galaxies enter
the cluster environment, most of there stars are tidally removed and a diffuse
spheroidal remnant remains (Moore \etal 1999). 
Numerical simulations of this process can be used
to study the diffuse cluster light component.

\

\centerline{\epsfysize=3.0truein \epsfbox{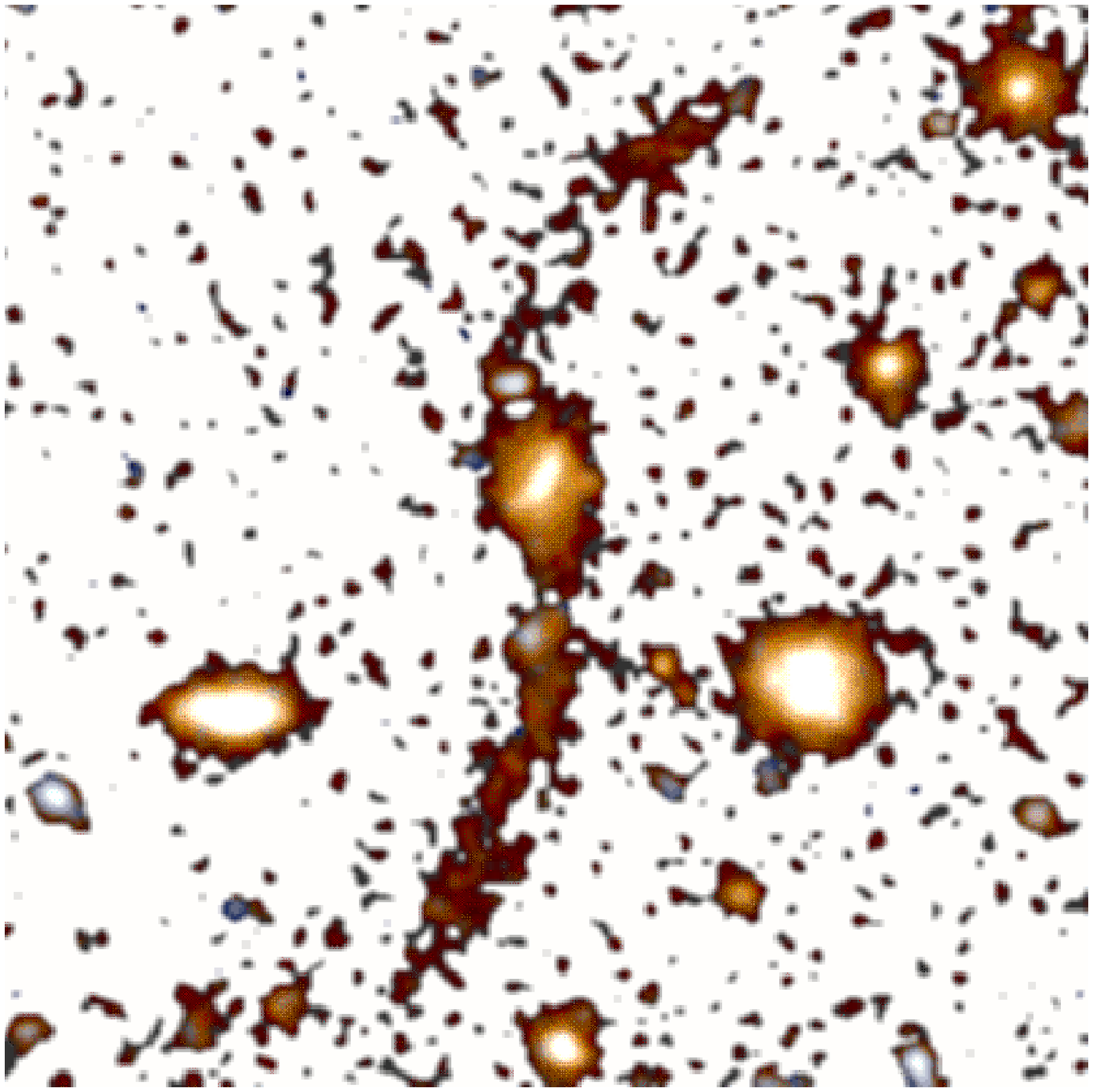}}

\

\noindent{\small {\bf Figure 2} \ \ An example of tidal tails from the
galaxy harassment process from the cluster CL0054-27 at z=0.56 courtesy of Ian Smail.
The image is roughly 150 kpc on a side.
Features this prominent are rare at this surface brightness ($\sim 27 M_{_B}$ 
per arcsec$^{-2}$) but are expected to surround all cluster dE (dSph) galaxies
at lower surface brightness levels.}




\section{Conclusions}

Two mechanisms are necessary to reproduce the observed morphological
evolution of galaxies in clusters.  At the faint end of the luminosity function,
disk galaxies are easily transformed into dE (dSph) galaxies by galaxy
harassment. Numerical simulations of this process predict that all dwarf
galaxies in nearby clusters will be embedded within very diffuse streams of
stripped stars.  Low surface brightness disks with slowly rising rotation curves
and shallow central potentials will not survive the fluctuating potentials of
rich clusters and will be tidally shredded to form the bulk of the diffuse
intra-cluster light.  Spirals with bulges are more stable to gravitational
encounters. They would retain their stars and gas and continue to form stars to
the present day. Most, if not all of the atomic hydrogen of these galaxies must
be rapidly removed upon entering the cluster environment in order to explain
their star-formation histories.  Fluid dynamic simulations of realistic disks
moving through a hot ICM demonstrates that a combination of ram-pressure and
viscous stripping can remove 100\% of the HI from the IGM within $10^7$
years. Continued tidal heating by encounters will complete a
transformation to the S0 class.

\end{document}